\documentclass[]{spie}  

\usepackage{amsmath,amsfonts,amssymb}
\usepackage{graphicx}
\usepackage[colorlinks=true, allcolors=blue]{hyperref}

\title{Correlation of interface transmission in THz spintronic emitters with spin mixing conductance in spin pumping experiments}

\author[a,*]{Evangelos Th. Papaioannou}
\author[b]{Laura Scheuer}
\author[b]{Moritz Ruhwedel}
\author[c]{Garik Torosyan}
\author[b]{Ren\'{e}  Beigang}
\author[a]{Georg Schmidt}

\authorinfo{Send correspondence to Evangelos Papaioannou, E-mail: evangelos.papaioannou@physik.uni-halle.de}

\affil[a]{Institute of Physics, Martin-Luther University Halle-Wittenberg, 06120 Halle, Germany}
\affil[b]{Department of Physics and Research Center OPTIMAS, University of Kaiserslautern, Kaiserslautern 67663, Germany}
\affil[c]{Photonic Center Kaiserslautern, Kaiserslautern 67663, Germany}

\authorinfo{$^{*}$Send correspondence to Evangelos Papaioannou, E-mail: evangelos.papaioannou@physik.uni-halle.de}

\begin{document} 
\maketitle

\begin{abstract}

The field of THz spintronics is a novel direction in the research field of spintronics that combines magnetism with optical physics and ultrafast photonics. The experimental scheme of the field involves the use of femtosecond laser pulses to trigger ultrafast spin and charge dynamics in bilayers composed of ferromagnetic (FM) and non-magnetic (NM) thin films where the NM layer features a strong spin-orbit coupling. The key technological and scientific challenges of THz spintronic emitters is to increase their intensity and frequency bandwidth. To achieve this the control of the source of the radiation, namely the transport of the ultrafast spin current is required. However, the transfer of a spin current from a FM to a NM layer is a highly interface-sensitive effect.
In this work we study the properties of the spin current transport through the interface measuring the strength of the THz emission and compare it to the effective spin mixing conductance, one of the key concepts in the spin current transport through interfaces. The results show an enhancement of the spin mixing conductance for interfaces with higher degree of epitaxy similarly to the improvement of the THz emission. The proportionality between spin mixing conductance and THz emission can define new directions in  engineering the emission of spintronic THz emitters.

\end{abstract}


\section{INTRODUCTION}
\label{sec:intro}  

Terahertz (THz) radiation (defined in the frequency range from 100 GHz to 30 THz), has long been studied in fields such as astronomy and analytical science. Recent technological innovations in optics and photonics enable THz research and technology to address an increasingly wide variety of applications: information and communications technology; biology, medical and pharmaceutical sciences; non-destructive evaluation homeland security; global environmental monitoring; THz sensor networks, and ultrafast computing. Despite the great advances over the last years, the field lacks of strong and broadband THz sources. 

Recent developments in nanomagnetism and spintronics and the first usage of ultrafast spin physics for THz emission has the potential to revolutionize the THz field and its applications\cite{Walowski:2016bt,Guo:20,Ding2020}. The physical mechanism of the THz radiation is based on heterostructures that consist of ferromagnetic (FM) and non-magnetic (NM) thin films. When triggered by ultrafast femtosecond (fs) laser pulses, they generate pulsed terahertz (THz) electromagnetic radiation due to the inverse spin Hall effect (ISHE), a mechanism that converts the spin currents originating in the magnetized FM layer into transient transverse charge currents in the NM layer resulting in THz emission~\cite{Kampfrath2013}. 

This new source of THz radiation is  an emerging topic open for intensive research. The efficiency of such emitters is in some cases  comparable with established type of THz sources (for example with nonlinear crystals)\cite{Seifert2016}. The technological and scientific challenge to engineer the emission is the main target of the research today. In this direction different strategies have been followed in order to explore the THz amplitude and bandwidth of the signal: different material compositions of FM/NM systems with a variety of thicknesses~\cite{Papa2019,Torosyan2018,Seifert2016,ADMA:ADMA201603031,ADOM:ADOM201600270,Papaioannou2018,Qiu:18}, ferri- and antiferromagnetic metal/Pt structures~\cite{spin2017,Albrecht2018,Ogasawara_2020}, spintronic emitters assisted by a metal-dielectric photonic crystal~\cite{Haifeng2018}, metallic trilayer structures with different patterned structures, interface materials and substrates~\cite{doi:10.1002/pssr.201900057,Li_2018,Seifert_2018,Li_2019,Garik2020,Hibberd2019,Kong2019} and THz emission using different excitation wavelengths~\cite{Papaioannou2018,Herapath2019}. 

However, even though the transfer of a spin current from a FM to a NM layer (that is the source of THz emission) is a highly interface-sensitive effect, only few works have tried to  correlate the structural quality of the interface with the THz signal strength and spectrum~\cite{Sasaki,Papa2019}. Recently, we have revealed that the performance of spintronic terahertz emitters can be controlled by optimizing the interface quality and its defect density\cite{Papa2019}. In particular, the presence of defect density results in changing the elastic electron-defect scattering lifetime in the FM  and NM layers and the interface transmission for spin-polarized, non-equilibrium electrons. A decreased defect density increases the electron-defect scattering lifetime and this results in a significant enhancement of the THz-signal amplitude and shifts the spectrum towards lower THz frequencies\cite{Papa2019}. Furthermore, besides the defect density the presence of the parameter of the interface transmission plays an important role. The latter is correlated to the ability of the interface to transfer hot carriers into the NM layer. It was shown\cite{Papa2019} that the interface transmission influences the spectral amplitude of the emitted THz field but conserves the composition of the spectrum.

However, the interface transmission is difficult to be directly addressed experimentally. In this work we address the interface transmission by measuring and comparing the THz emission with the effective spin mixing conductance in epitaxial Fe/Pt bilayer films. The effective spin mixing conductance is one of the key concepts in the spin current transport through interfaces when the excitation is performed in the GHz regime.  The results show that by changing the growth temperature the interface can be modified which in turn changes the spin mixing conductance and similarly the THz emission.  The proportionality between spin mixing conductance and THz emission can define new directions in  engineering the emission of spintronic THz emitters.

\section{Experimental results}
\subsection{Growth and structural analysis}

\begin{figure} [ht]
   \begin{center}
   \begin{tabular}{c} 
   \includegraphics[height=6cm]{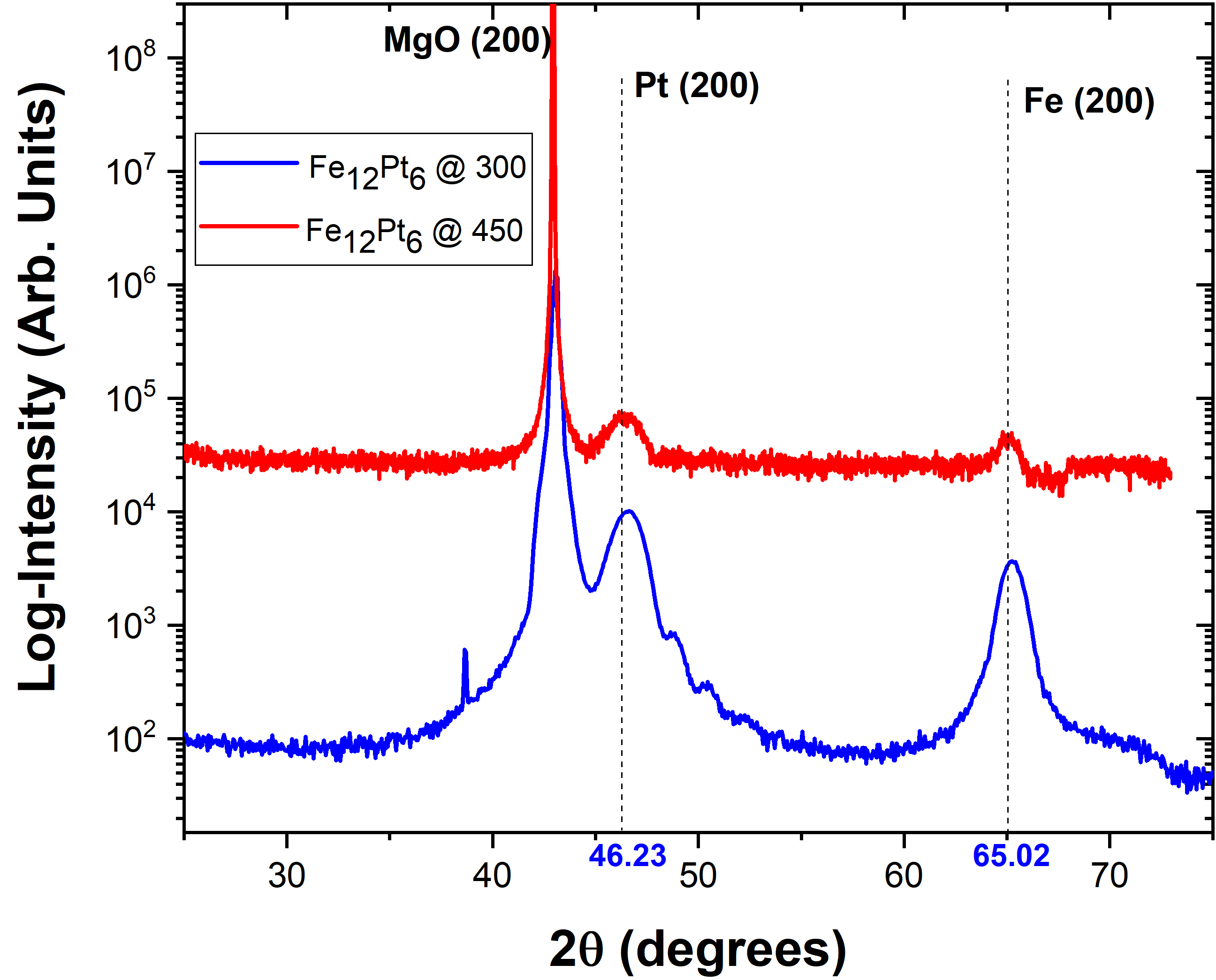}
   \end{tabular}
   \end{center}
   \caption[example] 
   { \label{fig:xrd} 
X-rays diffraction diagram of Fe (12 nm)/ Pt (6 nm) grown at 300$^\circ$C and at 450 $^\circ$C. The curves have been shifted for clarity.}
   \end{figure}

Fe/Pt bilayers were grown  on MgO (100) substrates by electron-beam evaporation in an ultrahigh vacuum (UHV) chamber with a base pressure of 5 $\times$ 10$^{-9}$ mbar. The growth rate was R = 0.05 $\text{\AA}$/s controlled by a quartz crystal during the deposition procedure. The incident Fe beam was perpendicular to the MgO substrate. The cleaning protocol of the MgO (001) 1 $\times$ 1 cm$^{2}$ substrates involved chemical cleaning with aceton and isospropanol and annealing at 150 \,$^\circ$C. The deposition of Fe was performed at 300 \,$^\circ$C substrate temperature. This growth temperature has been proven to provide the best Fe growth on MgO in our ultra-high vacuum chamber\cite{Keller2018}. After the deposition of Fe, annealing at the growth temperature was performed.
Subsequently a Pt layer was deposited on top of the Fe film. For the different samples different respective growth temperatures of room temperature, 150\,$^\circ$C, 300 \,$^\circ$C and 450\,$^\circ$C. Layer thicknesses were kept constant: Fe (12 nm)/ Pt (6 nm) and they were monitored in-situ by a calibrated quartz crystal oscillator and confirmed ex-situ by X-ray reflectivity (XRR) measurements. The aim was to keep the thickness the same and to change with the temperature the interface quality between Fe/Pt and so the interface transparency. Figure \ref{fig:xrd} presents X-ray diffraction (XRD) patterns of the Fe/Pt samples for different growth temperatures.  In all samples, peaks from the (200) crystal planes of the MgO substrate emerged at 2$\theta$ = 43$^{\circ}$  and the related Fe (200) peaks at approximately 65$^{\circ}$ are observed. Moreover, the Pt (200) peak emerges near 46 $^{\circ}$. The obtain values of both Fe and Pt are  very close to theoretical peak positions indicative of strain free growth along the growth direction. The preliminary structural characterization confirms that Fe/Pt layers grow epitaxially on MgO\,(100) substrates following the Bain epitaxial orientation that correlates the growth of an  \textit{fcc} lattice (Pt) on top of a \textit{bcc} lattice (Fe). The Fe lattice is in-plane rotated by  45$^\circ$ with respect to both MgO and Pt providing an epitaxial growth of the three lattices. 
\begin{table}[ht]
\caption{Summary of Fe/Pt samples grown at different temperatures} 
\label{tab:fonts}
\begin{center}       
\begin{tabular}{|l|l|l|l|} 
\hline
\rule[-1ex]{0pt}{3.5ex} $t_{\rm Fe}$ (nm) & Growth $T_{\rm Fe}$($^\circ$C) &  $t_{\rm Pt}$ (nm)& Growth $T_{\rm Pt}$ ($^\circ$C)\\
\hline
\rule[-1ex]{0pt}{3.5ex} 12 & 300$^\circ$C & 6 & 30 $^\circ$C  \\
\hline
\rule[-1ex]{0pt}{3.5ex}  12 & 300$^\circ$C & 6 & 150 $^\circ$C  \\
\hline
\rule[-1ex]{0pt}{3.5ex}  12 & 300$^\circ$C & 6 & 300 $^\circ$C   \\
\hline
\rule[-1ex]{0pt}{3.5ex}  12 & 300$^\circ$C & 6 & 450 $^\circ$C   \\
\hline
\end{tabular}
\end{center}
\end{table}

\subsection{Ferromagnetic resonance measurements}

In bilayer systems formed by a ferromagnetic (FM) layer in contact with a metallic nonmagnetic (NM) one, a pure spin current can be generated and injected into the NM layer when the ferromagnetic resonance (FMR) is excited. This process is commonly referred to as spin pumping\cite{RevModPhys.77.1375}.
The dynamic properties of the Fe/Pt samples were studied by measuring FMR using a strip line and a vector network analyzer (VNA-FMR). The dependence of the resonance linewidth $\Delta$H on the frequency $f_{\rm fmr}$  is used to determine the Gilbert damping parameter $\mathrm{\alpha}$: 

\begin{equation}
\mu_{0} \Delta H = \mu_{0} \Delta H_{0} + \frac{ 2 \alpha f_\text{fmr}}{\gamma }
\end{equation} 
\noindent with the inhomogeneous broadening $\Delta H_{0}$ being related to the film quality. 

\begin{figure} [ht]
   \begin{center}
   \begin{tabular}{c} 
   \includegraphics[height=6cm]{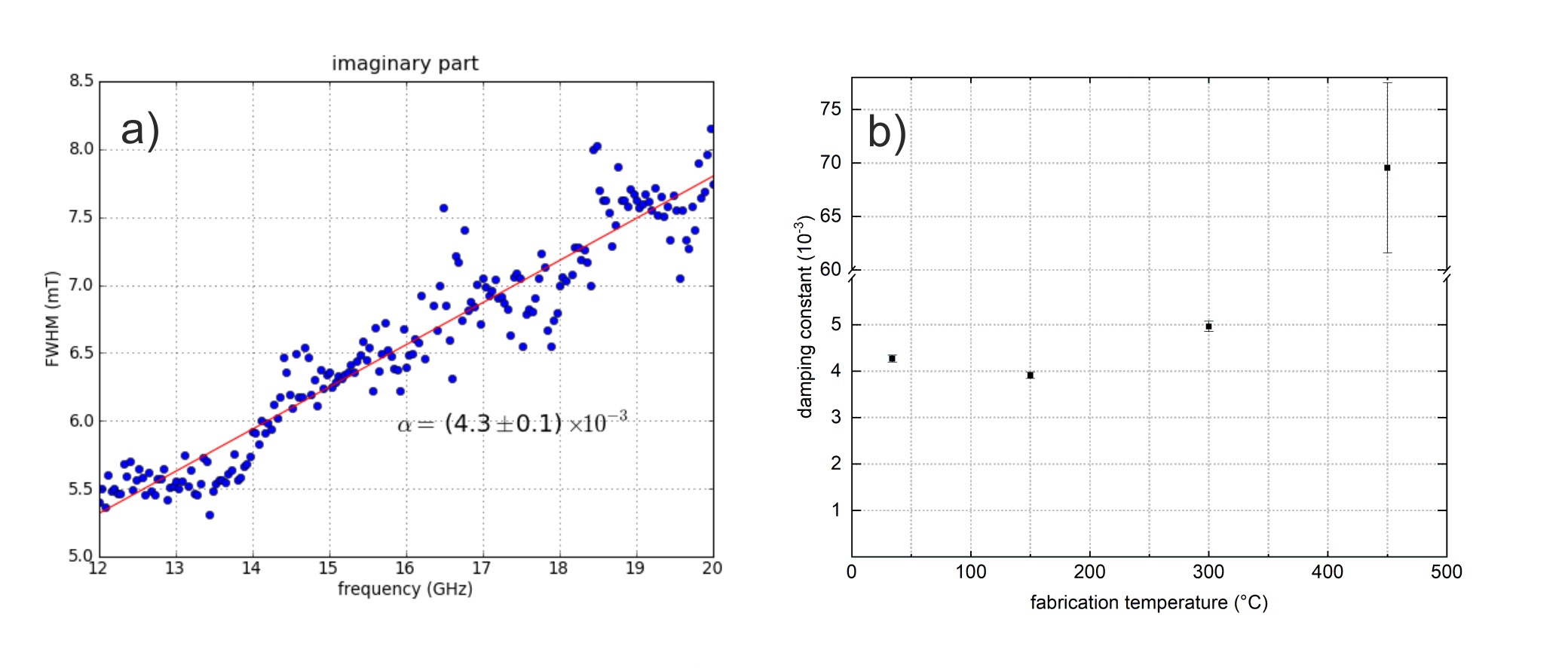}
	\end{tabular}
	\end{center}
   \caption[example] 
   { \label{fig:damping} 
a) Determination of the Gilbert damping from the dependence of the resonance linewidth $\Delta$H on the frequency $f_{\rm fmr}$. Here the data are shown for the sample with the Pt layer grown at 30 $^\circ$C sample. b) Gilbert damping parameter $\mathrm{\alpha}$ for different growth temperatures.} 
   \end{figure}

A typical example of the experimental determination of the Gilbert damping parameter $\mathrm{\alpha}$ from the measured linewidth  is given in Fig.\ref{fig:damping} a). The data refer to the sample with Pt-layer grown at room temperature. We calculated $\mathrm{\alpha}$ for all Fe/Pt samples shown in Table 1 and compare it to Fe film of the same thickness that was used as a reference sample. Since the spin current leaving the magnetic layer carries away angular momentum from the magnetization precession, it represents an additional loss channel for the magnetic system and consequently causes an increase in the measured Gilbert damping parameter.

The additional contribution of the spin current dissipation due to spin pumping is related to the spin mixing conductance:\cite{RevModPhys.77.1375}

\begin{equation} \label{conductance}
\Delta\alpha_{\rm sp}= \frac{\gamma\hbar}{4\pi M_\mathsf{s} \:d_{\rm FM}}g^{\uparrow\downarrow}_{\rm eff}.
\end{equation}

where $g^{\uparrow\downarrow}_{\rm eff}$  is the effective spin mixing conductance, which is controlling the magnitude of the generated spin current, $\gamma$ is the gyromagnetic ratio, $M_\mathsf{s}$ the saturation magnetization of each sample, $d_{\rm FM}$the thickness of the Fe layer.

The theoretical description\cite{RevModPhys.77.1375} of the spin mixing conductance $g^{\uparrow\downarrow}$  only includes the transmission of the interface which is determined by the materials and the interface quality. For the spin pumping also materials properties can become important which are then included in the so called $g^{\uparrow\downarrow}_{\rm eff}$\cite{Christoph2020}  which is determined by the measurement. In the case of Fe/Pt we will mostly observe the influence of $g^{\uparrow\downarrow}_{\rm eff}$ on the interface quality. It should be noted that the damping can also be influenced by effects like magnetic proximity or two magnon scattering\cite{Andres2ms}  which may complicate the analysis. 

The scope of these measurements is to  measure $g^{\uparrow\downarrow}_{\rm eff}$ and so to quantify the spin transmission through the interface between the ferromagnet (Fe) and the adjacent non-magnet (Pt).  The damping values for all samples are illustrated in in Fig.\ref{fig:damping} b). The damping is increasing with the growth temperature and so is the spin pumping efficiency. However for the sample grown at 450 $^\circ$C there is a dramatic increase to the damping that might not be only related to the spin pumping but also other effects can play a role like for example the two magnon scattering\cite{Andres2ms} from strongly modified interfaces.

\subsection{THz measurements}

\begin{figure} [ht]
   \begin{center}
   \begin{tabular}{c} 
   \includegraphics[height=4cm]{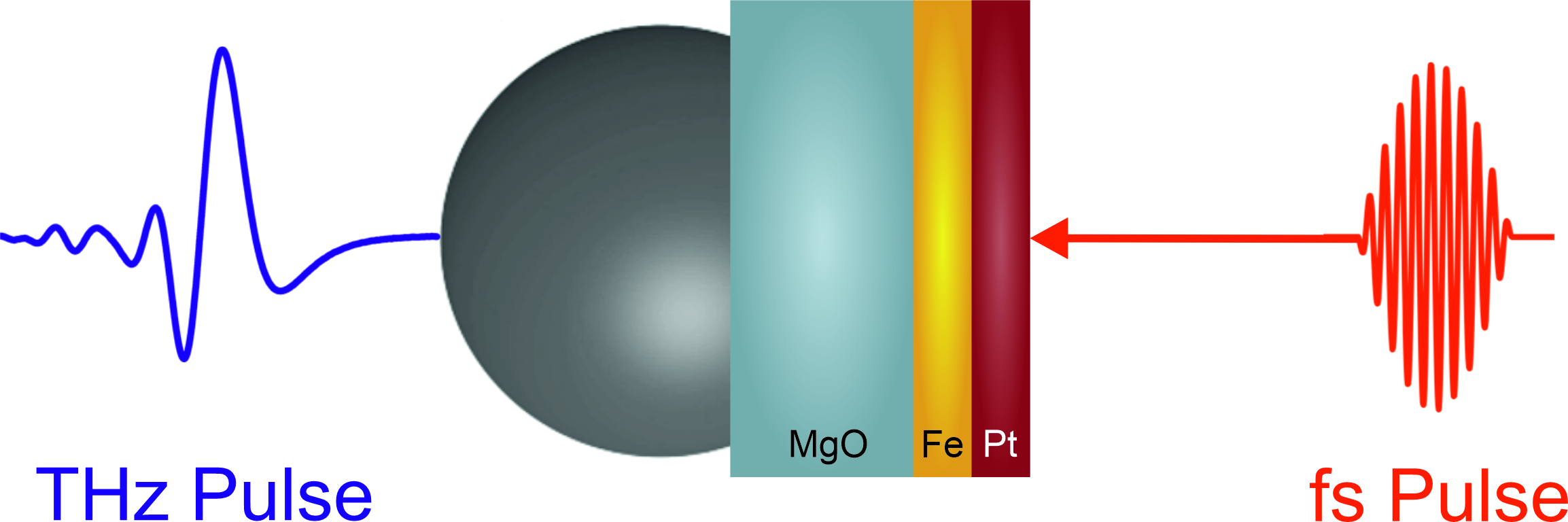}
   \end{tabular}
   \end{center}
   \caption[example] 
   { \label{fig:sample} 
Schematic of the THz measurement geometry. The fs laser pulse meets the sample from the Pt side and the THz pulse travels through the metal layers and the substrate and is collimated by the attached Si-lens.}
 
 \end{figure}

We turn now our attention to the estimation of the spin current transmission through the Fe/Pt inteface in the THz regime. The THz experiments with the Fe/Pt heterostructures were performed with a standard terahertz time domain spectroscopy (THz-TDS) system, where the heterostructures were used as THz emitters (the performance of the system is described in detail in Ref.[\citenum{Torosyan2018}]. The core of the system is a femtosecond Ti:Sa laser which produces 70\,fs optical pulses at a wavelength of 800\,nm with a repetition rate of 75\,MHz and a typical output power of 500\,mW. 
The probe beam is used to excite a photoconductive antenna (PCA) with a dipole length of 20\,\text{$\mu$}m that acts as THz detector. The spintronic emitter is magnetized by an external magnetic field (20\,mT) in the magnetic easy axis direction, which is perpendicular to the direction of the incident pump beam. The external field determines the polarization plane of the generated THz wave. The optical pump beam is focused onto the heterostructure by an aspherical short-focus lens. The Fe/Pt bilayer emits THz pulses into the free space as a strongly divergent beam. The pump beam is focused onto the emitter from the Pt side and a hyperhemispherical Si-lens is attached to the substrate of the emitter to collimate the beam, (Fig.~\ref{fig:sample}). The so formed conical THz beam is led via THz optics
to the PCA detector. To guarantee comparable experimental conditions, the alignment of the THz optics and of the detector is not changed during the exchange of the spintronic emitters. Since the lateral layer structure of the heterostructures is homogeneous and the position of the pump beam focus stays constant, the exchange of emitters does not influence the THz-signal. With the delay stage, the arrival of the THz pulse and the probe beam pulse is synchronized and the detected voltage at the PCA can be scanned. This voltage is proportional to the momentary electric field amplitude of the THz wave and, therefore, the THz-E-field and its phase can be measured as a function of the delay time. The voltage is measured by lock-in amplification while the pump beam is optically chopped. The THz spectral amplitude can be obtained by Fourier analysis. The bandwidth of the PCA detector with the 20 $\mu$m dipole length is limited to a minimum frequency of 100\,GHz and a maximum frequency of 8 THz. While the lowest measurable frequency is only limited by the dipole length of the PCA (longer dipole metalizations allow for the detection of lower frequencies), the detection of higher frequencies is limited by the strong phonon resonances of the GaAs substrate material of the PCA (absorption of the THz radiation).

\begin{figure} [ht]
   \begin{center}
   \begin{tabular}{c} 
   \includegraphics[height=7cm]{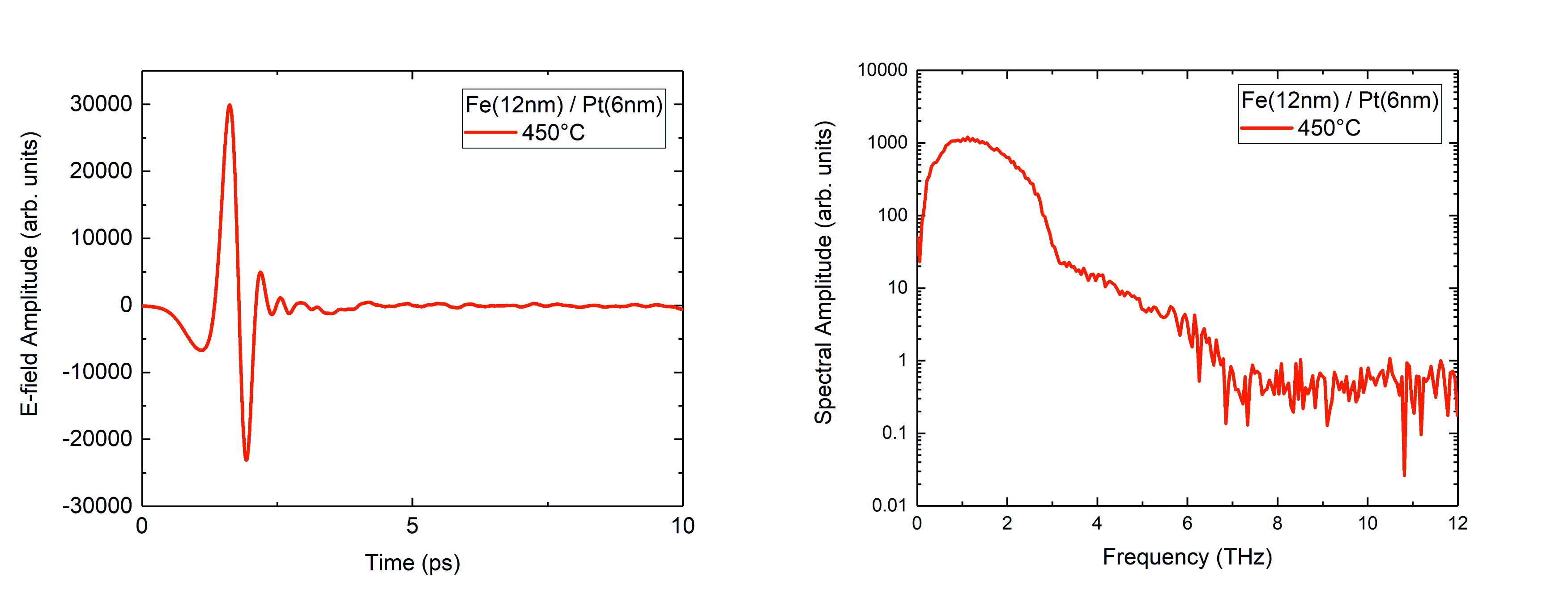}
   \end{tabular}
   \end{center}
   \caption[example] 
   { \label{fig:THzsignal} 
Experimental THz-E-field amplitudes of the Fe (12 nm)/Pt (6 nm) emitters grown at 450$^\circ$C on MgO in the time domain (left) and  the corresponding spectra (right). }
\end{figure}

\begin{figure} [ht]
   \begin{center}
   \begin{tabular}{c} 
   \includegraphics[height=6cm]{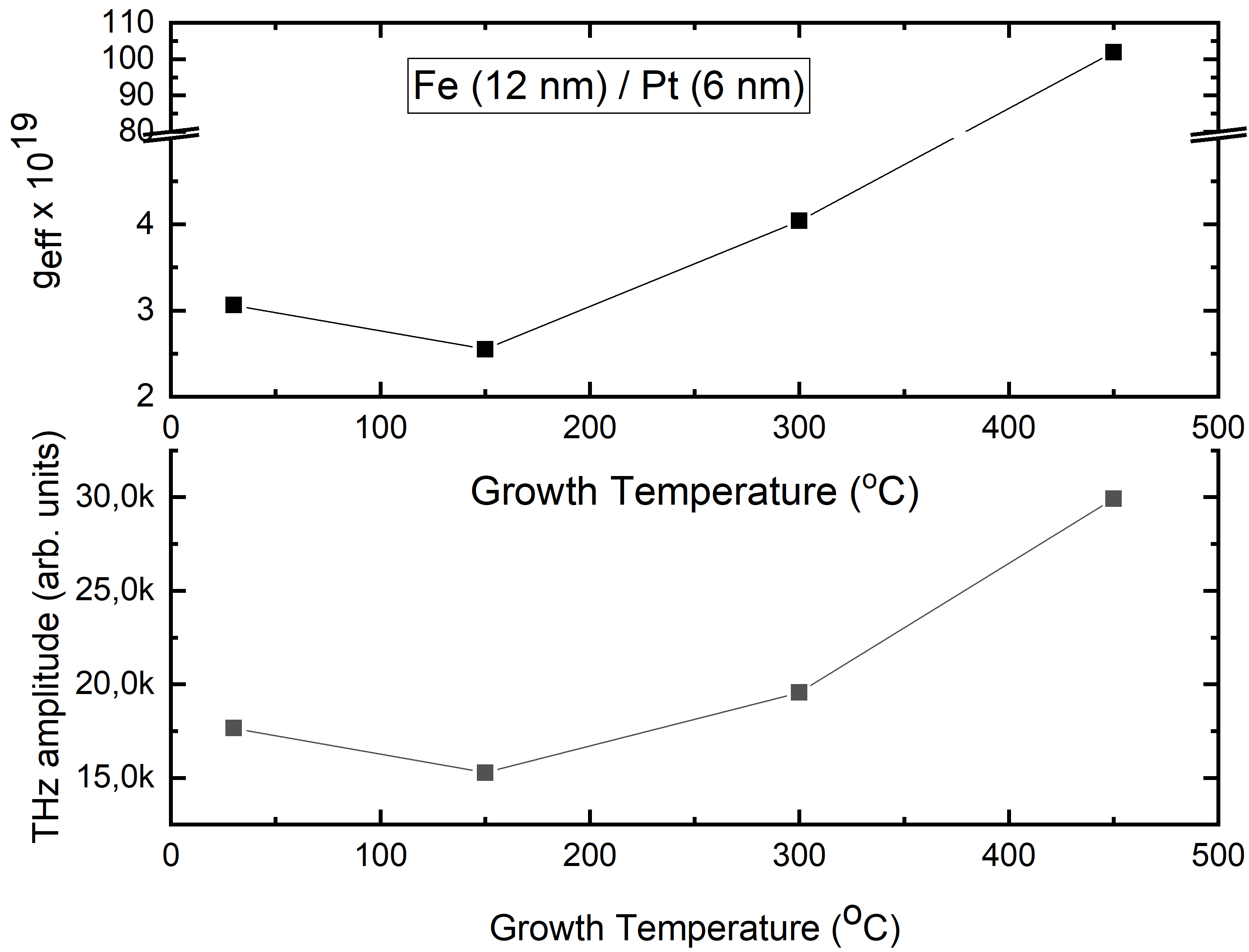}
   \end{tabular}
   \end{center}
   \caption[example] 
   { \label{fig:comparison} 
Upper panel: Effective spin mixing conductance of Fe (12 nm) / Pt (6 nm) samples series grown at different temperatures as indicated in the x-axis. The results are obtained from  Eq. \ref{conductance}.
Lower panel: THz amplitude with respect to the sample growth temperature. The amplitude of the THz emission is increased with temperature. The highest signal is observed for the Fe (12 nm)/Pt (6 nm) emitters grown at 450$^\circ$C.}
 
 \end{figure}

An example of a detected THz-signal and its corresponding spectrum is shown in Fig.\ref{fig:THzsignal}. The spectra are shown as measured and are not corrected for the detector response. Above 3 THz the well known strong THz absorption of MgO is visible for the MgO substrate.  Furthermore, the absorption around 8 THz in GaAs which is used as the semiconductor material for the detector antenna limits the bandwidth. The use of different detector (e. g. using electro-optical sampling in GaP) and even shorter pump pulses would lead to a much broader bandwidth up to 30 THz for these samples\cite{Papa2019,Seifert2016}. The maximum frequency measured in these experiments is determined by the frequency response of the dipole antenna of our photoconductive switch which was used as the detector and in our case is 1\,THz for all of our samples. All of the samples exhibit similar pulse shapes and bandwidths. The only difference is in the amplitude of the THz-field. The higher amplitude is measured for the 450$^\circ$C, see Fig.\ref{fig:comparison} lower panel. 
 In order to fully explain the evolution in the signal magnitude, the Fe/Pt interface transmission $T$ for the hot carriers has to be taken into account. A reduced transmission is not expected to alter the shape of the spectra but the amplitude of the THz signal will decrease gradually with lower T\cite{Papa2019}. Therefore, even a moderate deviation from almost perfect transparency is still  visible in our experiments. Although $T$ is an energy- and material-dependent quantity, we  regard it here as an independent parameter in our simplified discussion. Since the shape of the spectra remains the same for all the samples then the  electron scattering lifetimes (due to structural changes expressed in the defect density) should remain the same for all Fe/Pt samples\cite{Papa2019}.

In order to compare the THz interface transmission parameter $T$ with the quantifier  of the spin transmission through the interface between Fe and Pt in FMR measurements the effective spin mixing conductance $g^{\uparrow\downarrow}_{\rm eff}$  is plotted with temperature in upper panel of Fig.\ref{fig:comparison}. $g^{\uparrow\downarrow}_{\rm eff}$ follows the damping behaviour and increases with the growth temperature following the trend of the THz measurements.

In conclusion, we have grown Fe/Pt samples with same thicknesses. The films exhibit similar XRD patterns showing a good degree of epitaxy. In spite of the epitaxial relationship between Fe and Pt for all samples the interface morphology is different that leads to different transmission functions for the spin current as the enhancement of the effective spin mixing conductance and the THz reveals. We attribute this enhancement to the better interface transmission T for sample grown at higher temperatures. The physical mechanisms that give rise to the proportionality of effective spin mixing conductance and THz emission, although the excitation of spin current is performed in different time- and energy scales, is an open question that can define a new way to manipulate THz emission.

\acknowledgments 
We acknowledge financial support by the Deutsche Forschungsgemeinschaft (DFG) in the collaborative ressearch center TRR227, project B02.


\begin{thebibliography}{10}

\bibitem{Walowski:2016bt}
Walowski, J. and M{\"u}nzenberg, M., ``{Perspective: Ultrafast magnetism and
  THz spintronics},'' {\em Journal of Applied Physics}~{\bf 120},  140901--17
  (Oct. 2016).

\bibitem{Guo:20}
Guo, F., Pandey, C., Wang, C., Nie, T., Wen, L., Zhao, W., Miao, J., Wang, L.,
  and Wu, X., ``Generation of highly efficient terahertz radiation in
  ferromagnetic heterostructures and its application in spintronic terahertz
  emission microscopy (stem),'' {\em OSA Continuum}~{\bf 3},  893--902 (Apr
  2020).

\bibitem{Ding2020}
Chen, S.-C., Feng, Z., Li, J., Tan, W., Du, L.-H., Cai, J., Ma, Y., He, K.,
  Ding, H., Zhai, Z.-H., Li, Z.-R., Qiu, C.-W., Zhang, X.-C., and Zhu, L.-G.,
  ``Ghost spintronic thz-emitter-array microscope,'' {\em Light: Science \&
  Applications}~{\bf 9}(1),  99 (2020).

\bibitem{Kampfrath2013}
Kampfrath, T., Battiato, M., Maldonado, P., Eilers, G., N{\"o}tzold, J.,
  M{\"a}hrlein, S., Zbarsky, V., Freimuth, F., Mokrousov, Y., Bl{\"u}gel, S.,
  Wolf, M., Radu, I., Oppeneer, P.~M., and M{\"u}nzenberg, M., ``Terahertz spin
  current pulses controlled by magnetic heterostructures,'' {\em Nature
  Nanotechnology}~{\bf 8},  256 (03 2013).

\bibitem{Seifert2016}
Seifert, T., Jaiswal, S., Martens, U., Hannegan, J.and~Braun, L., Maldonado,
  P., Freimuth, F., Kronenberg, A., Henrizi, J., Radu, I., Beaurepaire, E.,
  Mokrousov, Y., Oppeneer, P.~M., Jourdan, M., Jakob, G., Turchinovich, D.,
  Hayden, L.~M., Wolf, M., M{\"u}nzenberg, M., Kl{\"a}ui, M., and Kampfrath,
  T., ``Efficient metallic spintronic emitters of ultrabroadband terahertz
  radiation,'' {\em Nat. Photon.}~{\bf 10},  483--488 (2016).

\bibitem{Papa2019}
Nenno, D.~M., Scheuer, L., Sokoluk, D., Keller, S., Torosyan, G., Brodyanski,
  A., L{\"o}sch, J., Battiato, M., Rahm, M., Binder, R.~H., Schneider, H.~C.,
  Beigang, R., and Papaioannou, E.~T., ``Modification of spintronic terahertz
  emitter performance through defect engineering,'' {\em Scientific
  Reports}~{\bf 9}(1),  13348 (2019).

\bibitem{Torosyan2018}
{G. Torosyan, S. Keller, L. Scheuer, R. Beigang, and E. Th. Papaioannou},
  ``Optimized spintronic terahertz emitters based on epitaxial grown fe/pt
  layer structures,'' {\em Sci. Rep.}~{\bf 8},  1311 (2018).

\bibitem{ADMA:ADMA201603031}
Wu, Y., Elyasi, M., Qiu, X., Chen, M., Liu, Y., Ke, L., and Yang, H.,
  ``High-performance thz emitters based on ferromagnetic/nonmagnetic
  heterostructures,'' {\em Advanced Materials}~{\bf 29}(4),  1603031 (2017).

\bibitem{ADOM:ADOM201600270}
Yang, D., Liang, J., Zhou, C., Sun, L., Zheng, R., Luo, S., Wu, Y., and Qi, J.,
  ``Powerful and tunable thz emitters based on the fe/pt magnetic
  heterostructure,'' {\em Advanced Optical Materials}~{\bf 4}(12),  1944--1949
  (2016).

\bibitem{Papaioannou2018}
Papaioannou, E.~T., Torosyan, G., Keller, S., Scheuer, L., Battiato, M.,
  Mag-Usara, V.~K., huillier, J.~L., Tani, M., and Beigang, R., ``Efficient
  terahertz generation using fe/pt spintronic emitters pumped at different
  wavelengths,'' {\em IEEE Transactions on Magnetics}~{\bf 54},  1--5 (Nov
  2018).

\bibitem{Qiu:18}
Qiu, H.~S., Kato, K., Hirota, K., Sarukura, N., Yoshimura, M., and Nakajima,
  M., ``Layer thickness dependence of the terahertz emission based on spin
  current in ferromagnetic heterostructures,'' {\em Opt. Express}~{\bf 26},
  15247--15254 (Jun 2018).

\bibitem{spin2017}
Seifert, T., Martens, U., Gunther, S., Schoen, M. A.~W., Radu, F., Chen, X.~Z.,
  Lucas, I., Ramos, R., Aguirre, M.~H., Algarabel, P.~A., Anadan, A., Karner,
  H.~S., Walowski, J., Back, C., Ibarra, M.~R., Morellan, L., Saitoh, E., Wolf,
  M., Song, C., Uchida, K., M{\"u}nzenberg, M., Radu, I., and Kampfrath, T.,
  ``Terahertz spin currents and inverse spin hall effect in thin-film
  heterostructures containing complex magnetic compounds,'' {\em SPIN}~{\bf
  07}(03),  1740010 (2017).

\bibitem{Albrecht2018}
Schneider, R., Fix, M., Heming, R., Michaelis~de Vasconcellos, S., Albrecht,
  M., and Bratschitsch, R., ``Magnetic-field-dependent thz emission of
  spintronic tbfe/pt layers,'' {\em ACS Photonics}~{\bf 5},  3936--3942 (10
  2018).

\bibitem{Ogasawara_2020}
Ogasawara, Y., Sasaki, Y., Iihama, S., Kamimaki, A., Suzuki, K.~Z., and
  Mizukami, S., ``Laser-induced terahertz emission from layered synthetic
  magnets,'' {\em Applied Physics Express}~{\bf 13},  063001 (apr 2020).

\bibitem{Haifeng2018}
Feng, Z., Yu, R., Zhou, Y., Lu, H., Tan, W., Deng, H., Liu, Q., Zhai, Z., Zhu,
  L., Cai, J., Miao, B., and Ding, H., ``Highly efficient spintronic terahertz
  emitter enabled by metal dielectric photonic crystal,'' {\em Advanced Optical
  Materials}~{\bf 0}(0),  1800965 (2018).

\bibitem{doi:10.1002/pssr.201900057}
Jin, Z., Zhang, S., Zhu, W., Li, Q., Zhang, W., Zhang, Z., Lou, S., Dai, Y.,
  Lin, X., Ma, G., and Yao, J., ``Terahertz radiation modulated by confinement
  of picosecond current based on patterned ferromagnetic heterostructures,''
  {\em physica status solidi (RRL) Rapid Research Letters}~{\bf 13}(9),
  1900057 (2019).

\bibitem{Li_2018}
Li, G., Mikhaylovskiy, R.~V., Grishunin, K.~A., Costa, J.~D., Rasing, T., and
  Kimel, A.~V., ``Laser induced {THz} emission from femtosecond photocurrents
  in {Co}/{ZnO}/{Pt} and {Co}/{Cu}/{Pt} multilayers,'' {\em Journal of Physics
  D: Applied Physics}~{\bf 51},  134001 (mar 2018).

\bibitem{Seifert_2018}
Seifert, T.~S., Tran, N.~M., Gueckstock, O., Rouzegar, S.~M., Nadvornik, L.,
  Jaiswal, S., Jakob, G., Temnov, V.~V., Munzenberg, M., Wolf, M., Klaui, M.,
  and Kampfrath, T., ``Terahertz spectroscopy for all-optical spintronic
  characterization of the spin-hall-effect metals pt, w and cu80ir20,'' {\em
  Journal of Physics D: Applied Physics}~{\bf 51},  364003 (aug 2018).

\bibitem{Li_2019}
Li, G., Medapalli, R., Mikhaylovskiy, R.~V., Spada, F.~E., Rasing, T.,
  Fullerton, E.~E., and Kimel, A.~V., ``Thz emission from co/pt bilayers with
  varied roughness, crystal structure, and interface intermixing,'' {\em Phys.
  Rev. Materials}~{\bf 3},  084415 (Aug 2019).

\bibitem{Garik2020}
{Mag-usara}, V.~K., {Torosyan}, G., {Talara}, M., {Afalla}, J., {Muldera}, J.,
  {Kitahara}, H., {Scheuer}, L., {Sokoluk}, D., {Papaioannou}, E.~T., {Rahm},
  M., {Beigang}, R., and {Tani}, M., ``Spintronic thz generation using a
  silicon-based fe/pt bilayer as the radiation source,'' in [{\em 2019 44th
  International Conference on Infrared, Millimeter, and Terahertz Waves
  (IRMMW-THz)}{\nolinebreak\hspace{0.1em}]},   1--2 (2019).

\bibitem{Hibberd2019}
Hibberd, M.~T., Lake, D.~S., Johansson, N. A.~B., Thomson, T., Jamison, S.~P.,
  and Graham, D.~M., ``Magnetic-field tailoring of the terahertz polarization
  emitted from a spintronic source,'' {\em Applied Physics Letters}~{\bf
  114}(3),  031101 (2019).

\bibitem{Kong2019}
Kong, D., Wu, X., Wang, B., Nie, T., Xiao, M., Pandey, C., Gao, Y., Wen, L.,
  Zhao, W., Ruan, C., Miao, J., Li, Y., and Wang, L., ``Broadband spintronic
  terahertz emitter with magnetic-field manipulated polarizations,'' {\em
  Advanced Optical Materials}~{\bf 7}(20),  1900487 (2019).

\bibitem{Herapath2019}
Herapath, R.~I., Hornett, S.~M., Seifert, T.~S., Jakob, G., Kl{\"a}ui, M.,
  Bertolotti, J., Kampfrath, T., and Hendry, E., ``Impact of pump wavelength on
  terahertz emission of a cavity-enhanced spintronic trilayer,'' {\em Applied
  Physics Letters}~{\bf 114}(4),  041107 (2019).

\bibitem{Sasaki}
{Y. Sasaki, K.Z. Suzuki, and S. Mizukami}, ``Annealing effect on laser
  pulse-induced thz wave emission in ta/cofeb/mgo films,'' {\em Appl. Phys.
  Lett.}~{\bf 111},  102401 (2017).

\bibitem{Keller2018}
{S. Keller, L. Mihalceanu, M. R. Schweizer, P. Lang, B. Heinz, M. Geilen, T.
  Br{\"a}cher, P. Pirro, T. Meyer, A. Conca, D. Karfaridis, G. Vourlias, T.
  Kehagias, B. Hillebrands, and E. Th. Papaioannou}, ``Determination of the
  spin hall angle in single-crystalline pt films from spin pumping
  experiments,'' {\em New J. Phys.}~{\bf 20},  053002 (2018).

\bibitem{RevModPhys.77.1375}
Tserkovnyak, Y., Brataas, A., Bauer, G. E.~W., and Halperin, B.~I., ``Nonlocal
  magnetization dynamics in ferromagnetic heterostructures,'' {\em Rev. Mod.
  Phys.}~{\bf 77},  1375--1421 (Dec 2005).

\bibitem{Christoph2020}
Hauser, C., Ballani, C., Durrenfeld, P., Heyroth, F., Trempler, P., Ebbinghaus,
  S.~G., Papaioannou, E.~T., and Schmidt, G., ``Enhancement of spin mixing
  conductance in la0.7sr0.3mno3/lanio3/srruo3 heterostructures,'' {\em physica
  status solidi (b)}~{\bf n/a}(n/a),  1900606 (2020).

\bibitem{Andres2ms}
Conca, A., Keller, S., Schweizer, M.~R., Papaioannou, E.~T., and Hillebrands,
  B., ``Separation of the two-magnon scattering contribution to damping for the
  determination of the spin mixing conductance,'' {\em Phys. Rev. B}~{\bf 98},
  214439 (Dec 2018).

\end{thebibliography}

\end{document}